# Physics discovery in nanoplasmonic systems *via* autonomous experiments in Scanning Transmission Electron Microscopy

Kevin Roccapriore,[1] Sergei V. Kalinin,[2,a] and Maxim Ziatdinov[1,3,b]

[1] Center for Nanophase Materials Sciences, Oak Ridge National Laboratory, Oak Ridge, TN, USA, 37831

[2] Department of Materials Science and Engineering, University of Tennessee, Knoxville, TN, USA, 37916

[3] Computational Sciences and Engineering Division, Oak Ridge National Laboratory, Oak Ridge, TN, USA, 37831

Physics-driven discovery in an autonomous experiment has emerged as a dream application of machine learning in physical sciences. Here we develop and experimentally implement a deep kernel learning workflow combining the correlative prediction of the target functional response and its uncertainty from the structure, and physics-based selection of acquisition function, which autonomously guides the navigation of the image space. Compared to classical Bayesian optimization methods, this approach allows to capture the complex spatial features present in the images of realistic materials, and dynamically learn structure-property relationships. In combination with the flexible scalarizer function that allows to ascribe the degree of physical interest to predicted spectra, this enables physical discovery in automated experiment. Here, this approach is illustrated for nanoplasmonic studies of nanoparticles and experimentally implemented in a truly autonomous fashion for bulk- and edge plasmon discovery in MnPS$_3$, a lesser-known beam-sensitive layered 2D material. This approach is universal, can be directly used as-is with any specimen, and is expected to be applicable to any probe-based microscopic techniques including other STEM modalities, Scanning Probe Microscopies, chemical, and optical imaging.

[a] sergei2@utk.edu
[b] ziatdinovma@ornl.gov

The emergence of advanced electron and scanning probe microscopies has opened fundamentally new opportunities for exploring the physics of nanoscale and atomic systems.[1–3] Over the last decades, multiple advances ranging from mapping structure of ferroic[4,5] and charge density wave systems,[6,7] to exploring the nature of electronic and superconductive order parameters in quantum materials,[8,9] to probing vibrational and plasmonic properties on the nanometer and atomic level[10–12] has become commonplace. The advent of monochromated and aberration-corrected electron microscopy[13] combined with the beam engineering[14] and low-temperature environments[15] opens the pathway towards further advancements.

In scanning transmission electron microscopy (STEM), one of the most prominent recent experimental directions is the high-energy resolution low-loss electron energy loss spectroscopy (EELS) for probing phononic and plasmonic properties.[16,17] These responses contain the information on the local dielectric function convolved with local geometry,[18–20] and hence contain the key to understanding the physics of collective excitations in the nanoscale systems. Similarly, plasmons strongly couple to local chemical reactivity and form the basis for chemical characterization or even beam-controlled chemical reactions. Finally, the development of multiple classes of quantum devices hinges on the capability to understand and harness phonon and plasmon functionalities on the nanometer levels.

These considerations have stimulated the large body of experimental effort in EEL spectroscopy and EELS imaging of 2D materials, nano-optical systems, semiconductors, and many more.[21,22,13] Currently, the experimental studies in STEM-EELS, much like virtually all other areas of modern physical imaging, are based on the human operator paradigm. The regions for spectral imaging, whether single point spectroscopy or a grid of points, are selected based on operator intuition, and data analysis is performed after (in some cases, long after) it has been acquired. At the same time, generative theoretical models for many of these are not well understood, have a large number of poorly known parameters, or are absent. Hence, experimental discovery of new physical relationships and behaviors is often limited by chance, i.e., whether the object or behaviors of interest happened to be observed during the experiment and if they contain features identifiable by an operator and collaborating team. Note that multiple calls for storing full experimental data streams and making the data open that are now prevalent in multiple scientific communities[23,24] seek to address the second aspect.

These factors have launched much interest in automated and autonomous experiment as an approach to accelerate the rate of physical discovery, both in the context of electron microscopy, scanning probe microscopy methods, as well as scattering techniques. Over the last several years, multiple opinion pieces towards development of automated experiment (AE) in microscopy have been published.[25,26] However, the targets of the automated experiment have not been generally analyzed in detail, with vast majority of applications to date reporting either probe-based manipulation based on rigid discovery and manipulation rules,[27,28] implementation of deep neural network-based computer vision algorithms towards identification of objects of interest over large scan regions,[29–31] or dynamic sampling routines determined in the feature space.[32–34] At the same time, automated discovery of novel physical phenomena has remained elusive. Currently, the primary bottleneck towards implementation of the autonomous experiment in microscopy is the lack of an algorithm for active learning targeting *physical* discovery.

Indeed, in many areas of scientific research including X-Ray scattering, automated synthesis, and recently SPM and STEM, the primary paradigm for the AE is Gaussian Processes (GP) and GP-based Bayesian optimization (BO).[35–38] However, the extant implementations of BO use only the information acquired during the experiment to select the new locations for exploration. Similarly, the relationship between the explored locations is defined by the functional form of the kernel function (somewhat equivalent to the parametrized correlation function discovered during the experiment), thus offering only limited flexibility. While new forms of kernel functions are becoming available and the non-parametric and analytic forms of acquisition functions can be balanced, the lack of prior knowledge utilization is common to the BO methods that are also usually limited to low-dimensional parameter space. Secondly, the learning of the kernel structure is effectively equivalent to the discovery of the correlation function parametrized *via* a certain functional form, and hence their capability to represent complex non-periodic spatial structures prevalent in microscopic data is extremely limited. Structural discovery was recently explored in detail using BO,[39] but only insignificant improvements were realized compared to rectilinear grid scans.

At the same time, in the typical microscopy experiment, the locations for the detailed spectroscopic or imaging studies is selected based on specific features in the structural data set. For example, in nanoplasmonic structures we want to collect EEL spectra on "special" locations associated with certain positions with respect to geometric structures. This information is not available for conventional BO. Ideally, we would like to balance this feature-based exploration with the perceived value of the already-discovered behaviors, for example, intensities in a specific energy region, functional form of the spectrum, presence of features that are expected to be signatures of physical phenomena of interest, or exploration targeting the discovery of new behaviors (i.e., curiosity learning). It is also important to mention that very often scientists operating the microscope have a degree of anticipation – based on prior knowledge and experience, comparisons with other similar systems, or general physical considerations – on how the interesting behaviors are likely to manifest in the observed data. These considerations are often formulated in a somewhat flippant form as "we know interesting physics when we see it." However, this level of analysis until now was unavailable for autonomous microscopy.

Here, we use the combined power of BO and feedforward neural networks in the form of deep kernel learning (DKL) to enable the autonomous physical discovery in microscopy and implement it with STEM-EELS of nanoplasmonic structures. In this approach, we perform active learning of structure-property relationships as single-shot learning, rather than reconstruction of an image or spectrum (either data driven or using physical priors) from pre-acquired sparse data.[40–42] In other words, the algorithm actively interrogates the microscope during operation, rather than working with the data samples according to a certain pattern after it is acquired. We further develop an approach to incorporate physical knowledge in the discovery in the form of a scalarizer function that explicitly allows to define what physics is of interest. Note that the scalarizer can be even trained by human – and then the experiment will discover interesting physical signatures following the model set by a trained physicist. We develop the associated workflow, quantify its performance on ground truth data of plasmonic nanoparticles, and deploy it on the operational microscope such that it operates autonomously without any additional hardware to actively discover the edge plasmon functionality in MnPS$_3$,[43–46] a relatively unexplored beam-sensitive 2D antiferromagnet.

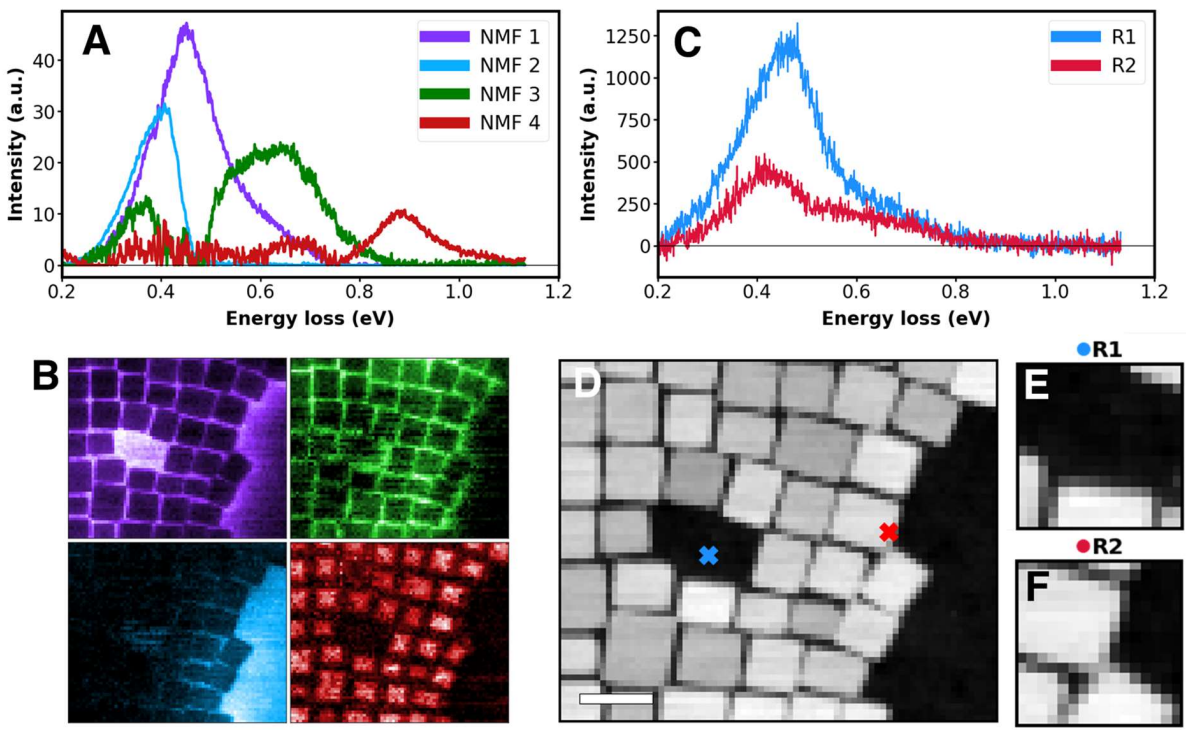

**Figure 1**. System overview and interpretation of data by processing techniques. NMF deconvolution for $N = 4$ components where spectral endmembers and abundance maps are shown in A and B, respectively. Sub-image formation from structural image in D is shown in E and F, where EEL spectra in C correspond to response from center pixel in sub-images E and F. The data set of the plasmonic cubes is used from [[47,48]]. Scale bar in D is 20 nm.

As a first model system, we have chosen fluorine and tin co-doped indium oxide nanoparticles whose plasmon response lies in the near infrared. This is the same material system studied in previous works[49,50,48,47,51] and we find it to be an excellent choice for visualizing the spatial dependence of a variety of plasmonic features that depend on geometry and local particle organization. The typical structural image obtained via high-angle annular dark field (HAADF) scattering and several spectra are shown in Figure 1. It is common to show deconvolved components using exploratory data techniques such as non-negative matrix factorization (NMF) for visualizing the locality of the spectral behavior.[52–55] Here, we show the NMF in the supplementary materials and code which is available in the notebook, but we do not discuss it here. Figure 1 (A and B) largely illustrate the rich plasmonic features possible in this system.

A crucial aspect of NMF and other spectrum-based linear and manifold learning methods is that they do not consider any relationships to the structural data or the spatial location from which the spectra are acquired. However, it is these relationships which guides the human operator to decide where to perform measurements. Previously, we have introduced the *im2spec* approach,[47,56] using the encoder-decoder neural network architecture to establish the correlative relationship between local structure and functional responses. However, the *im2spec* approach requires a full data set to perform the analysis. Furthermore, transferring a pretrained deterministic neural network to autonomous experiment is limited both by out-of-distribution effects and the capability of the network to "discover" only the phenomena it has been trained on. In other words, the approach where the first part of the experiment is used to train the network, and the second uses thus trained networks to explore (A/B testing[57]) are known to be least data efficient and most sensitive to the changes in the data generation process.

With this in mind, and using the same dataset, we illustrate the deep kernel learning approach, where the fully available structural data is used to actively navigate the spectrum "acquisition" process. Similar to the *im2spec* method, we split the structural image into a set of patches of size $w$ at every spatial coordinate and pair these with the EEL spectrum at that coordinate. The patches are hence the structural descriptors, whereas EEL spectra are the target functionalities (see Figure 2). Figure 1 C-F illustrates the patch formation and corresponding spectra. Here, we assume that the structural information is available everywhere, whereas spectra are available only sequentially as the algorithm actively interrogates the microscope to perform sequential spectrum acquisition. In this process, the subsequent locations in the image plane are chosen based on the results obtained from previous locations and predefined criteria that defines degree of interest to prediction.

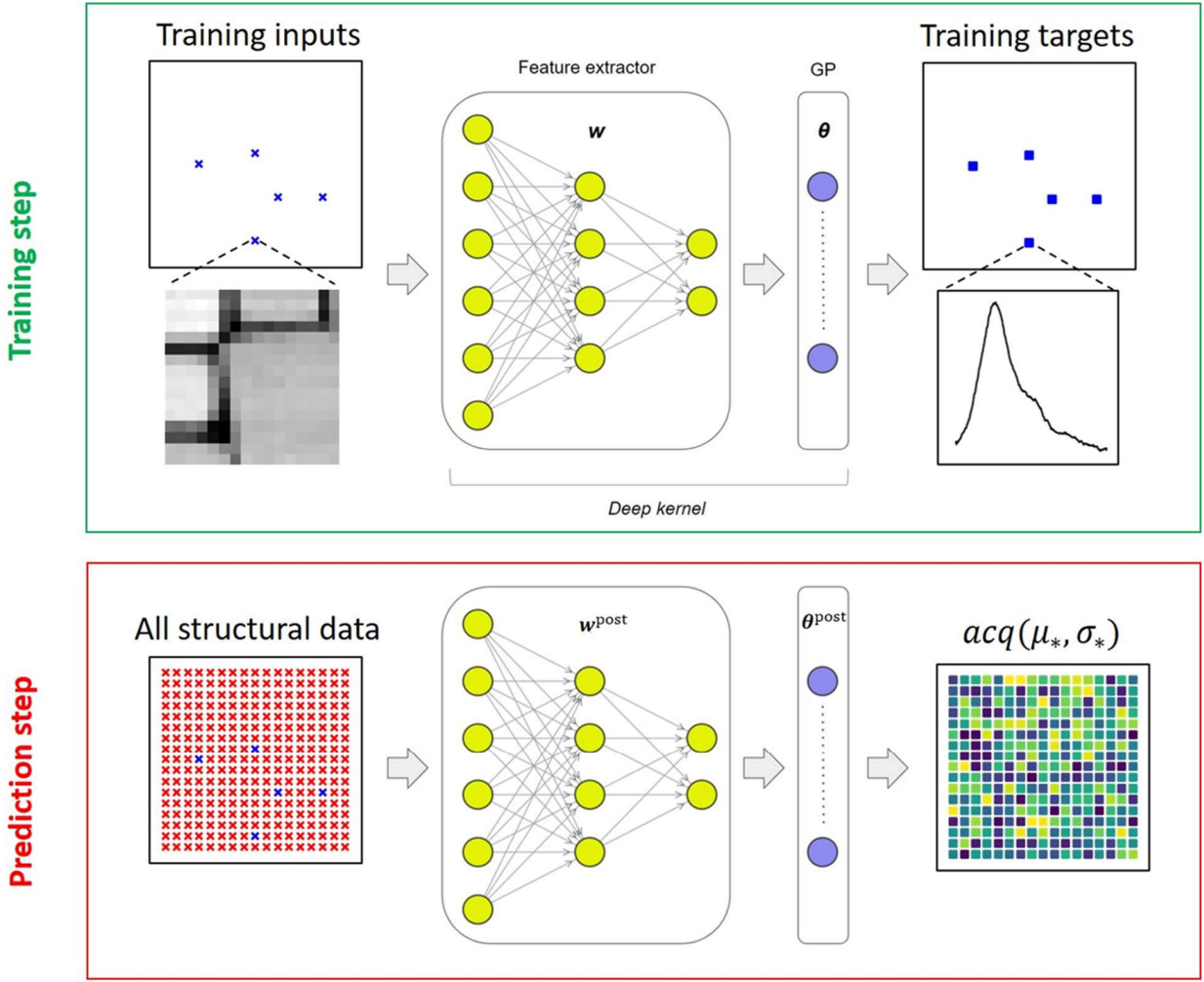

**Figure 2.** Schematic illustration of a single deep kernel learning (DKL) step for active learning of system properties where an input is a structural image, and an output is either a spectrum or a specific scalar property derived from it. Computationally wise, a feedforward neural network takes structural images and embeds them into a low-dimensional feature space in which a standard ('base') Gaussian process kernel operates. At training step, the weights ($w$) of the neural network and the hyperparameters ($\theta$) of the base kernel are optimized jointly using a small number of available measurements. Once a DKL model is trained, it can be used to compute expected function value ($\mu_*$) and the associated uncertainty ($\sigma_*$) for all structural data (for which there are no measured spectra). These are used to derive the acquisition function, $acq(\mu_*, \sigma_*)$, for selecting the next measurement point according to $x_{\text{next}} = \text{argmax}(acq)$.

To illustrate the behavior of DKL in this system, we develop several examples demonstrating its potential to reconstruct the data for full sampling, followed by sparse (random) sampling, and subsequently for active sampling. These examples hence cover a range of applications such as denoising and pansharpening, batch updates, and particularly active learning with physics-based discovery. Note that the post-acquisition tasks (pansharpening) can be also implemented with different algorithms such as compressed sensing and Gaussian Processes.[58–61] However, the DKL is crucial in enabling physics-based active learning and is demonstrated here. In the latter case, we assume that we have access to the full structural information within the image and the EEL spectra at several locations. We postulate the existence of a (unknown) relationship between local structure and EEL spectrum and seek to reconstruct it for the full data set. However, here we assume that only partial information on functionality is available.

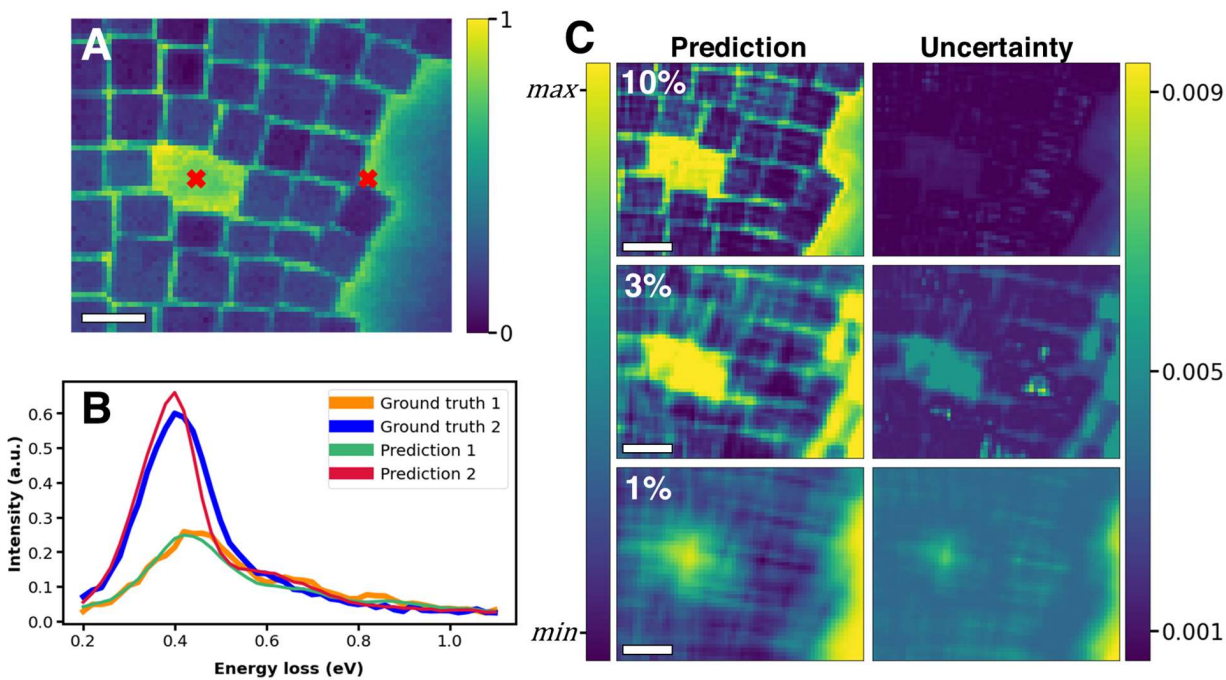

**Figure 3.** DKL model performance for full data reconstruction. (A) Ground truth image representing energy average over all energies (normalized), with selected ground truth spectra and DKL spectral predictions using all data shown in (B). DKL predictions and associated uncertainties using partial data (indicated by percentages) are shown in (C). 10%, 3%, and 1% partial data represent 280, 85, and 28 measurements, respectively. Predictions exist on relative scale. All scale bars 20 nm.

The DKL reconstruction of structure-property relationships for random selection of sampling points is shown in Figure 3. Individual spectral reconstruction using all known points for selected coordinates in A is shown in B, which are compared to the ground truth spectra. Supplemental materials contain additional spectral comparisons. We note that the DKL reconstructs the spectra from images. However, for visualization purposes we further calculate the energy-averaged images across the entire spectrum, which are shown in Figure 3C for 10%, 3%, and 1% fraction of known points, where the corresponding uncertainties in the spectral reconstructions are also shown. Even for only 1% known points, the reconstruction is strikingly good, indicating that utilizing structural similarity allows to harness a wealth of information contained in the structure-spectra relationships. It is important to consider here that the partial data is chosen randomly, and the model is trained all at once for each percentage set of partial data. In other words, this is not a scenario of active learning where the model is continuously retrained with each new measurement, and therefore random initializations may have an impact on the behavior of the model.

To rationalize these observations, we note that there is a relationship between local structures and EEL spectra. Generally, it can be expected that similar structures have similar spectra and discovering the relationship between the two constitutes the correlative model. The deviations from such relationships signify the presence of unknown mechanisms, with the uncertainty maps providing likely spatial locations on where they manifest. However, the points in this case are chosen randomly (i.e., absent any knowledge of the system) and hence behaviors of interest localized in a small number of spatial locations remain poorly explored.

The analysis above illustrates the power of DKL to reconstruct spectra from structure based on partial data and yield corresponding uncertainties after the experiment. However, the ultimate target of autonomous experiment is not only reconstruction of behaviors, but also the search for specific behaviors of interest. To understand fundamental principles behind this process, we note that each human-led experiment is guided by the combination of exploratory and exploitative strategies. Exploratory strategies generally target the discovery of possible behaviors in the system. In exploitative strategies, we seek to find new behaviors, refine the parameters of the known models, or refine a certain hypothesis.

The flexibility of the DKL allows exploration of the image space based on the combination of the predicted target function and its uncertainty, where both are vectors. This allows significant flexibility in defining exploration and exploitation strategies, based for example on the maximum value of a function, total integrated uncertainty, or other criteria. However, this also provides a unique opportunity to define physics-based acquisition functions that target the behaviors of interest as reflected in spectra. In other words, much like human-based exploration of data after acquisition ("we know interesting physics when we see it"), here we can formulate the exploration criterion based on the physical behavior of interest, whether it is specific peak intensity or width, peak intensity ratios, presence of certain features, similarity to previously known behaviors, etc.

Here, we first illustrate this approach on the model data, with the full spectral data set (invisible to the DKL algorithm) being available as a ground truth. As a test case, after removing the zero-loss peak (ZLP) in the usual manner[62,63] by fitting a power law to the ZLP, the maximum spectral intensity is used as the optimization parameter for DKL. For the plasmonic nanoparticles under test, it turns out that the maximum intensity generally corresponds to a collective plasmon resonance which is most strongly observed when the electron probe is near - but not on - particles. By merely attempting to maximize spectral intensity, in this case, DKL can determine the relationship between intense spectral response and local geometries which partially contain particles. In comparison to the cluster learning in Figure 3 where the structure-property relationship is learned from a given segment of the full data all at once, Figure 4 illustrates the discovery process at several different stages (number of measurement points), where the overlaid red points on the HAADF-STEM images in Figure 4C are the locations of where the EELS measurements are taken. A structure-property relationship is established with each new measurement, where the subsequent measurement location is deeply related to the previous accumulation of structure-property knowledge. Emphasis is placed on the fact that these four images actually show snapshots of the active learning pathway by where the measured points are acquired. Each EEL spectrum is converted into a scalar – in this case, the maximum spectral intensity after ZLP removal. This scalar is considered the measurement for DKL, and with each new measurement, DKL is retrained. Figure 4C illustrates this at four different stages throughout the active learning process, where the maximum valued pixel in the map of the acquisition function is the location of the next measurement. Strikingly, even at very few measured points, the model has learned the relationship between structure in the HAADF and maximizing plasmon intensity, as illustrated by the acquisition functions and model predictions in Figure 4C. If individual reconstructed spectra at different percentages of acquired points are compared to one another, in general, there is not a considerable difference between even 0.1% and 10% spectral predictions. This is shown in the Supplementary materials in more detail; however, the primary message remains the same: the spectral relationship to the local structure is determined very early in the experiment, and is only refined with more sampling points and training.

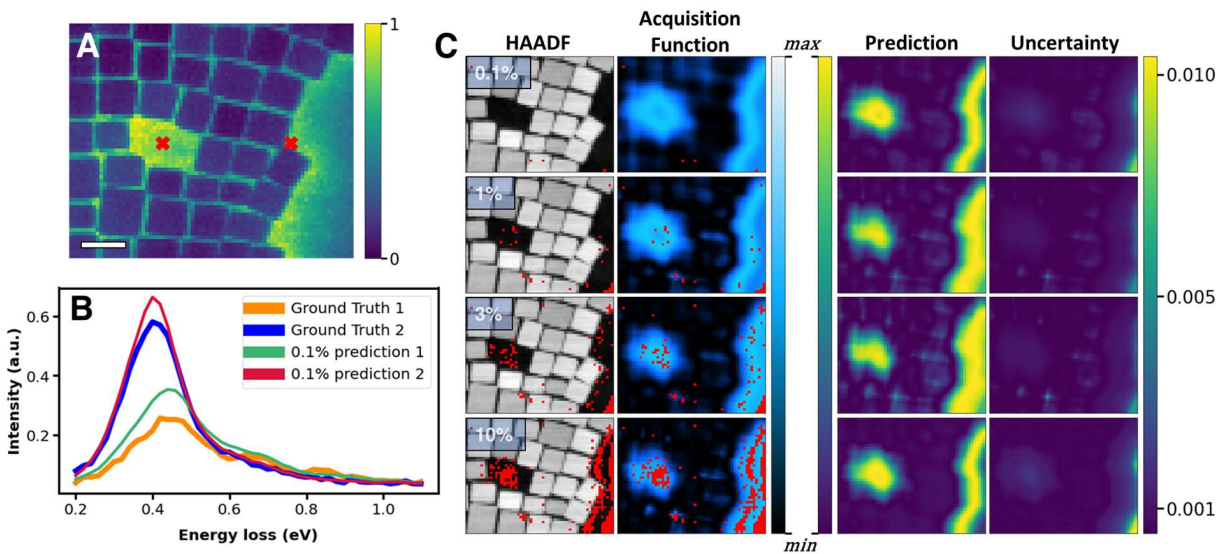

**Figure 4.** DKL discovery pathway for edge plasmon search. Normalized two-dimensional map of maximum EEL spectral intensity (after ZLP removal) is shown in A, where the model's goal is to optimize the peak near 0.4 eV in (B). Additional spectral comparisons are presented in Supplementary materials. DKL pathway shown in panel (C) for increasing number of measured points; sampled points are shown in HAADFs and (scalarized) acquisition functions in red, with DKL predictions and associated uncertainties also shown for where to best optimize maximization of the plasmon peak. Acquisition function and prediction maps are both relative quantities. Percentages in (C) indicate the total amount of points that have been measured and used in training up to that point, where 0.1%, 1%, 3%, and 10% represent 3, 28, 85, and 280 measurements, respectively. Scale bar in A is 20 nm.

Finally, we proceed to implement the DKL-based discovery in the autonomous experiment implemented on an operational Scanning Transmission Electron Microscope. Here, we used the NION Monochromated Aberration Corrected STEM (MACSTEM) and utilized the Python-scripting environment enabled by the SWIFT software suite[64] to establish the communication between DKL control routines and the beam position. A link between the EELS camera and SWIFT is already present since we use the NION IRIS spectrometer and camera system, therefore all data modalities (EELS, 4D-STEM, ADF) are accessible directly *via* a Python interface. Additional details are available in Materials and Methods section.

Here, we illustrate this approach for physics-driven discovery of plasmonic behavior discovery in $MnPS_3$, a less-known layered 2D van der Waals antiferromagnet. By analogy with other layered materials, we hypothesize that this material system should demonstrate specific plasmonic behaviors in the vicinity of interfaces, i.e., develop edge plasmons. The physical criterion for edge plasmon detection is the presence of the peak in EELS spectrum below the bulk plasmon. Hence, here we choose to optimize a peak ratio in the EEL spectrum where we posit the existence of a low energy electronic feature as well as a higher energy bulk plasmon resonance. The EEL spectra predicted from the structural descriptors are fit with two Lorentzian functions whose peak positions can fluctuate in energy but not overlap. The ratio of the amplitudes of the two are then used as the guiding principle in the autonomous experiment, where we seek to maximize the low energy feature and simultaneously minimize the bulk plasmon. Emphasis is placed on the fact that *a priori* we have no knowledge of any edge plasmon behavior in this material – we merely suggest a physical criterion in which we are interested. Note that this scalarizer function that defines the degree of physical interest can be considerably more complex and be realized as a decision tree, neural network trained on human-labeled data, or even reflect human decisions made during the experiment. In other words, it allows to incorporate a full spectrum of methods available to the well-trained human operator.

For this specific experiment, equally important is that this material is rather sensitive to electron beam irradiation, making high-fidelity hyperspectral imaging difficult. This lends well to the paradigm of autonomous experiment which allows to only irradiate regions that are deemed to be of interest, limiting the total dose the specimen receives. Lastly, it is crucial to convey the fact that the DKL algorithm has not been pre-trained and no sample information has been supplied to guide or otherwise assist in its performance. In other words, the following experiment applies to practically *any* never-before-seen specimen.

The autonomous experiment (AE) utilizing the DKL workflow is illustrated in Figure 5, where a HAADF-STEM and selected EEL spectra of an exfoliated 2D flake consisting of several layers are shown. We first demonstrate in Figure 5C the DKL pathway, prediction, and acquisition function as a function of number of measured points when we choose to optimize the peak ratio previously discussed. For clarity, the acquisition function effectively combines the knowledge of the DKL prediction with minimizing the uncertainty. Fairly quickly the DKL recognizes the relationship between the boundary of the flake and vacuum, in which a strong localized edge plasmon is found to exist and the higher energy bulk plasmon is weak or non-existent. Note that the top edge is more favorable than other edges presumably due to a sharper boundary. After measuring between 10 and 25 EEL spectra (which corresponds to only 1% of the entire space), it is apparent that the DKL predicts the location of edge modes rather well (Figure 5C, DKL Prediction). This has implication that after only a fraction of points are initially sampled, the model either need not be trained further or not as frequently, allowing a substantial speed increase.

Note that the choice of the physical model influences the discovery pathway and consequently the features that are discovered. Once again, this is similar to the human-based workflow, when depending on the specific hypothesis the operator will choose a different sequence of measurements. To illustrate this, we change the criterion from peak ratios to maximum peak intensity as is shown in Figure 5D. In this case, the maximum peak intensity appears to correspond to the bulk plasmon resonance, which generally increases in strength with thickness. As we would expect, since the intensity in the HAADF can be treated as a measurement of relative thickness (at these scales), we observe the DKL pathway to preponderantly explore the regions *on* the flake. This second search criterion serves to support the fact that this autonomous experiment enabled by DKL indeed does search for physics that depends on the built-in models that are provided. Comparing the two pathways in Figure 5C and 5D, a different number of acquired points are needed to satisfactorily learn the structure-property relationships. We selected to stop the experiments after acquisition of 100 points for both pathways only for demonstration and comparison purposes; in a practical setting, one might choose to halt the training after a threshold level of uncertainty has been reached and proceed to acquire additional points (without further training), which also allows dramatic increase in speed of the remaining experiment. In the supplementary information, videos are provided which show the algorithm in action with more detail.

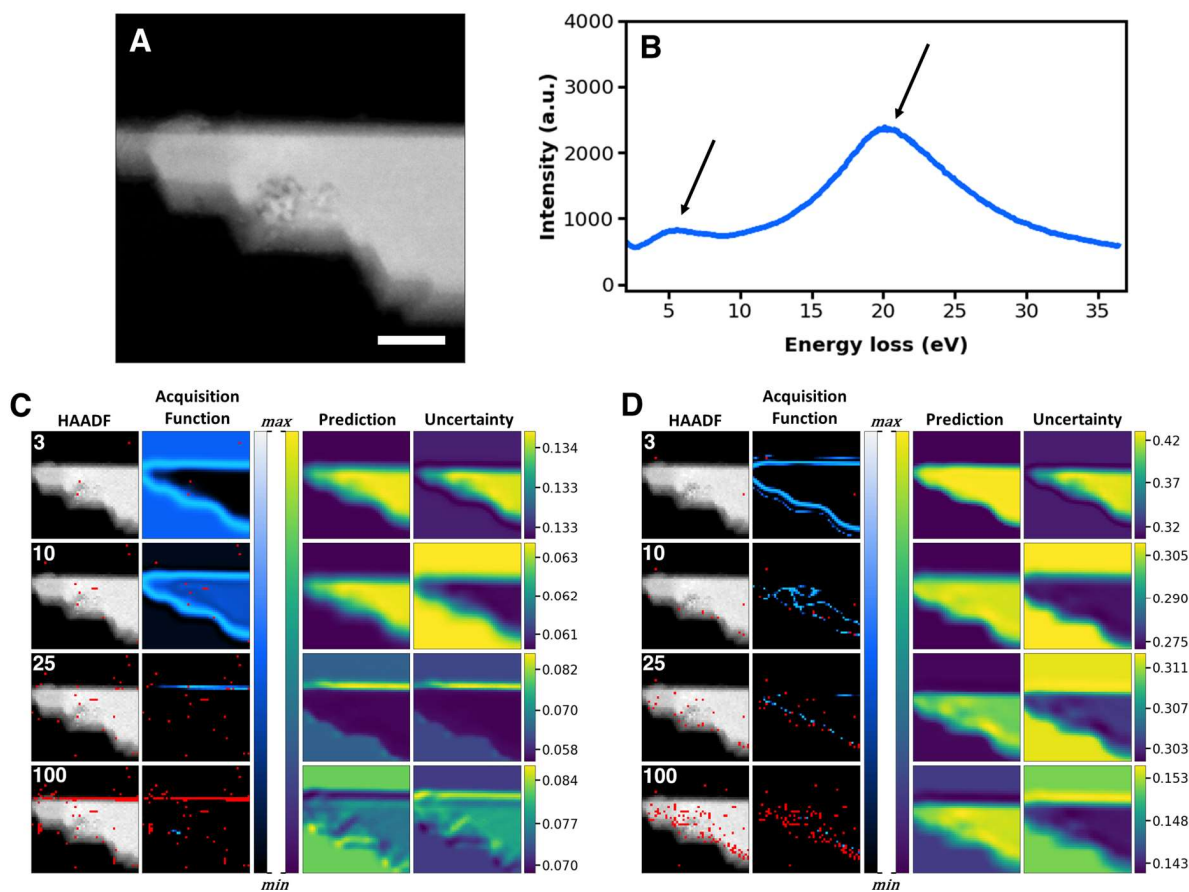

**Figure 5.** Autonomous experimentation (AE) enabled by DKL. Autonomous experiment with structural (HAADF) image of 2D section of suspended $MnPS_3$ shown in (A); autonomous peak ratio optimization shown in (C), while autonomous optimization of maximum spectral intensity shown in (D). Average spectrum from all visited points in autonomous experiment in (C) is shown in (B), where arrows indicate the peaks from which a peak ratio is determined. Maximum spectral intensity in most space tends to relate to high energy peak near 20 eV. Process shown for 3, 10, 25, and 100 acquired measurements for two different physics search criteria employed: peak ratio optimization (C) and maximization of spectral intensity (D). Scalebar in (A) 50 nm. Acquisition function and prediction maps are both relative quantities.

To summarize, we have implemented the physics discovery workflow for electron energy loss spectroscopy in scanning transmission electron microscopy. This approach uses the structural information as initial data for scientific discovery and uses active learning to build the structure-property relationships in the system under investigation. The discovery process is guided by the physics-based acquisition function that allows incorporation of physical models, target functionalities, or discovery of novel phenomena. Depending on experimental goals, this approach can provide a full hyperspectral reconstruction (so-called vector DKL), as is performed in Figures 3 and 4; or as in Figure 5, it can solely focus on the specified structure-property relationship (scalar DKL) to conduct autonomous microscopy. Note that one may choose to combine these and perform the full vector DKL with the autonomous experiment, but at the expense of slightly increased computation costs, although there is not necessarily a benefit to perform the full hyperspectral reconstruction during the experiment. Here, this approach enabled the detection and realization of edge plasmon behavior in the 2D layered antiferromagnet $MnPS_3$. No prior knowledge of edge plasmon activity has been mentioned in literature or was expected to occur, and while it is true that such a phenomenon could have been discovered without DKL – it was not – and precisely demonstrates the power of using DKL for physical discovery.

Having illustrated the active learning approach with DKL for $MnPS_3$, once again it is stressed that DKL is not limited to certain material systems – on the contrary, any material that can be placed in the microscope can be used without any prior knowledge of the material system. This active learning of structure-property relationships (with quantified uncertainties) that are valid for the observed material system is the first distinctive aspect of this approach. The second key point about the DKL discovery is that almost any physical criteria can be put forth – this may be as simple or as complicated as desired. In STEM, the autonomous experimentation can be readily extended beyond the spectral measurements such as EELS and EDS, towards 4D-STEM modes. Here, scalar-derived quantities such as lattice parameters, strain, and electric fields that are calculated from a diffraction pattern, along with their uncertainties, can be used as a part of the exploration criteria. With the myriad of different modalities that are possible in the electron microscope and other probe-based imaging such as variants of Scanning Probe Microscopy, DKL holds the key to opening the floodgates of autonomous experimentation.

Perhaps even more importantly, similar principles (sequential scanning of feature rich parameter spaces) underpin multiple other areas of scientific discovery, from instrumental characterization methods to the discovery of new materials, molecules, etc. Hence, we believe that the proposed principle combining the correlative ML prediction of complex responses combined with physics-based navigation can be of interest for multiple other areas of scientific research.

**Acknowledgement:** This effort (electron microscopy, feature extraction) is based upon work supported by the U.S. Department of Energy (DOE), Office of Science, Basic Energy Sciences (BES), Materials Sciences and Engineering Division (K.M.R., S.V.K.) and was performed and partially supported (M.Z.) at the Oak Ridge National Laboratory's Center for Nanophase Materials Sciences (CNMS), a U.S. Department of Energy, Office of Science User Facility. The authors acknowledge Shin-Hum Cho and Delia J. Milliron for supplying semiconducting nanoparticles as well as Nan Huang and David G. Mandrus for the $MnPS_3$ used in this work.

## Materials and methods

### Materials

F,Sn:In$_2$O$_3$ nanoparticles were chemically synthesized by Shin-Hum Cho and Delia J. Milliron using standard Schlenk line techniques with a modification of previously reported methods for continuous slow injection synthesis of indium oxide nanoparticles as described in [[50]].

MnPS$_3$ crystals were grown by Nan Huang and David G. Mandrus using a chemical vapor transport (CVT) method as described in [[65]]. Single crystals were mechanically exfoliated directly onto Au Quantifoil TEM grids.

### Scanning Transmission Electron Microscopy

All STEM experiments were carried out using the NION MACSTEM with an accelerating voltage of 60 kV and probe semi-convergence angle of 30 mrad. EEL spectra were acquired using 100 ms dwell time per pixel while HAADF-STEM images were acquired with 16 us per pixel. For the F,Sn:In$_2$O$_3$ nanoparticle system, the electron source was monochromated such that the full width at half max (FWHM) of the ZLP was approximately 40 meV, and a dispersion of 0.8 meV/ch (2048 channels) was selected. This resulted in a nominal probe current of 30 pA. With MnPS$_3$, the source was not monochromated due to the presence of higher energy excitations, giving a FWHM of 350 meV and nominal probe current of 200 pA, using a dispersion of 20 meV/ch.

### Autonomous Microscopy

As part of the NION Swift package, direct access *via* a Python scripting interface exists to both monitor and control various aspects of the NION MACSTEM. In our case, the electron probe position, EELS camera, and Ronchigram camera are all accessible with appropriate commands. This implies there is no additional hardware requirement for autonomous experiments using the NION systems. Installation of various required Python packages (e.g., PyTorch, SciPy, etc.) was done such that the model training is performed on the microscope hardware PC. The CPU or GPU can be used for training, where in the form of scalar DKL and small/sparse data, use of a GPU presents no significant benefit. For vector DKL (i.e., full spectrum prediction), use of GPU is practically required due to needed parallelization of many spectral targets.

Autonomous experiment begins with the collection of HAADF-STEM (structural) image of size $m \times n$ pixels, followed by $N$ number of randomly sampled EEL spectra in the same space, where here $m \times n$ is 50 x 50 and $N$ is usually two or three. The probe is then immediately moved to a pre-defined safe location (e.g., for beam sensitive samples) or is blanked. Image patches (features) are created for all pixels in the HAADF using a user-specified window size, e.g., 8 x 8 pixels, which should be appropriately selected based on field of view and structural details that may be contained in each image patch.

Searching for particular physical effects is facilitated by specification of search criteria – in scalar DKL, the EEL spectrum is reduced to a single scalar quantity in any imaginable way. Note that in vector DKL, the entire spectrum does not need to be reduced to a scalar quantity, though depending on GPU, the number of spectral channels may be limited. In this work, we showed examples of two relatively simple criteria with scalar DKL– maximum spectral intensity and a peak ratio of fitted curves. Curve fitting is performed by defining a Lorentzian function and using the SciPy package in Python.

The DKL model is trained with each additional EELS measurement such that it continuously learns structure-property relationships. This is adjustable to the point where training may be halted at a certain number of measurements, or a pre-trained model may be used, however the use of pre-trained models is not discussed here as our focus is that of *active* learning. A built-in exploration parameter also exists in the model which allows for the balance of exploitation and exploration of the data space, i.e., this might help to avoid remaining in a local minimum.

**Gaussian Process Regression and Bayesian Optimization**

In the Gaussian process regression setting, given inputs $X = (x_1, \ldots, x_N)$ and targets $\mathbf{y} = (y_1, \ldots, y_N)$, we assume that each target point is generated according to[66]

$$y_n = f(x_n) + \varepsilon_n, \tag{1}$$

where $f$ is drawn from a Gaussian process (GP) prior, $f \sim \mathcal{GP}(0, k)$, and $\varepsilon_i \sim \mathcal{N}(0, s_n^2)$. The covariance (kernel) function $k$ defines a strength of correlation between points in the input space and its hyperparameters are learned from the data (along with the noise variance) by performing a gradient ascent on the log marginal likelihood. The most common choice of kernel is a radial basis function (RBF) which on one hand supports a large class of functions with various shapes but on the other hand have inductive biases toward very simple solutions.

Once a GP model is trained (i.e., the kernel and noise variance are learned), it can be used to compute predictive mean ($\mu_*$) and variance ($\Sigma_*$) for a new (test) point $x_*$ as

$$\mu_* = \mathbf{k}_*^\top (K + s_n^2 I)^{-1} \mathbf{y}, \tag{2a}$$

$$\Sigma_* = k(\mathbf{x}_*, \mathbf{x}_*) - \mathbf{k}_*^\top (K + s_n^2 I)^{-1} \mathbf{k}_* \tag{2b}$$

where $K$ is the $n$-by-$n$ matrix of covariances computed for all pairs of training points and $\mathbf{k}_*$ is the vector of covariance between the test and training points. In the context of microscopy/spectroscopy, the training data ($X$ and $\mathbf{y}$) correspond to sparse measurements over a selected field of view (FOV) defined on rectangular grid of size $i \times j$ whereas the new/test points ($X_*$) correspond to yet unmeasured regions from the same FOV. The goal of Bayesian optimization (BO) is to discover regions where a particular functionality or structure is maximized using a minimal number of measurements. The next measurement points are selected using the maxima of so-called acquisition function(s) derived from the GP-predicted mean, $\mu_*$, and uncertainty, $\sigma_* = \text{diag}(\Sigma_*)$ on the test data. The popular choice of acquisition functions is the expected improvement (EI), which tells a likelihood of the highest improvement over the current "best measurement"[67] and is defined as

$$\alpha_{EI} = (\mu(\mathbf{x}) - y^+ - \xi)\Phi\left(\frac{\mu(\mathbf{x})-y^+-\xi}{\sigma(\mathbf{x})}\right) + \sigma(\mathbf{x})\phi\left(\frac{\mu(\mathbf{x})-y^+-\xi}{\sigma(\mathbf{x})}\right) \qquad (3)$$

where $\Phi$ is a standard normal cumulative distribution function, $y^+$ is the best predicted value, $\phi$ is the standard normal probability density function, and $\xi$ balances the exploration and exploitation (here it was set to 0.01).

Unfortunately, the classical GP has several major limitations that preclude its widespread adoption for real-time microscopy and spectroscopy measurements. First, it does not scale well to high-dimensional ($D \gtrsim 3$) inputs making it unsuitable for hyperspectral measurements ($D \geq 3$) in online settings. Second, the classical GP does not actually learn the representations from the data. As such, it does not allow for using information obtained through different experimental modalities for more efficient navigation during the BO. For example, it is not uncommon to see the associations between features observed in easy-to-acquire structural 2D images and those seen in spectroscopic data (the existence and strength of such associations depend on the particular system and experimental method), but it is not possible to exploit such associations in the BO setting with the classical GP approach.

**Deep Kernel Learning**

One of the solutions to the aforementioned limitations of the classical GP is to combine a deep neural network with a Gaussian process. The former is well-known to be a powerful deterministic tool for representation learning that can easily scale to high-dimensional inputs whereas the latter provides the reliable uncertainty estimates. This hybrid solution is known as deep kernel learning (DKL)[68] which we schematically illustrated in Fig 1. Its central idea is straightforward. The high-dimensional input data is embedded by a feed-forward neural network $g_\omega$ into the (latent) feature space where a standard ("base") GP kernel operates. The new effective kernel is expressed as $k_{DKL}(x, x') = k(g_\omega(x), g_\omega(x'))$, where $\omega$ are the parameters (weights and biases) of the neural network. The parameters of the base kernel and the weights of the neural network are trained jointly by maximizing the log marginal likelihood. Once trained, the DKL-GP model can make a prediction with quantified uncertainty on new (high-dimensional) input data for selecting the next measurement point according to Eq (2)-(3).

In the DKL GP, the inputs can be image patches extracted at ($k$, $l$) locations of the 2D grid and the targets can be either full spectra in those locations or some specific characteristic(s) of spectra such as intensity of selected peaks (note that the dimensionality of each input patch is equal to the total number of pixels contained in the patch; hence, even for a patch of $8 \times 8$ the input dimensionality is 64). In this case, the DKL GP will be learn a correlative structure-property relationship whereas the uncertainty in property prediction from patches of structural data at (*k*, *l*) points on the grid can be used for selecting the next measurement point using a pre-defined acquisition function. We note that while predicting property from structure can be also done with the *im2spec* type of encoder-decoder networks,[43,49] it does not output predictive uncertainty and hence cannot be used for the BO-based autonomous experimentation.

      We note that during the exact GP training, the entire dataset must be passed "through" a model at once, which may not be possible for modern neural network architectures containing millions and sometimes billions of parameters even when using state-of-art Graphics Processing Units (GPUs). One possible solution is to use a stochastic variational (i.e., approximate) inference that allows for mini-batching of training data. At the same time, we found that in many cases the experimental observations can be well modeled by simple few-layer perceptrons (MLPs) with a much smaller number of parameters that can accommodate an entire dataset allowing for the exact GP training and inference. We also note that it is possible to use a neural network (pre-)trained in the offline setting (e.g., using the previously collected data) and "freeze" its weights during the DKL training on new data from a similar experiment, which could save computational time/cost in the online setting.

**Data availability**

The data used for analysis as well as additional materials are available through the Jupyter notebook located at: https://github.com/kevinroccapriore/AE-DKL/